\documentclass[conference]{IEEEtran}
\IEEEoverridecommandlockouts
\usepackage{cite}
\usepackage{amsmath,amssymb,amsfonts}
\usepackage[noend]{algorithmic}
\usepackage{algorithm}
\usepackage{graphicx}
\usepackage{textcomp}
\usepackage{xcolor}
\usepackage{multirow}
\def\BibTeX{{\rm B\kern-.05em{\sc i\kern-.025em b}\kern-.08em
    T\kern-.1667em\lower.7ex\hbox{E}\kern-.125emX}}
\begin{document}


\title{Towards Efficient IMC Accelerator Design Through Joint Hardware-Workload Co-optimization}

 \author{\IEEEauthorblockN{Olga Krestinskaya${^1}$, Mohammed E. Fouda${^2}$, Ahmed Eltawil${^1}$, Khaled N. Salama$^1$}
 \IEEEauthorblockA{\textit{${^1}$King Abdullah University of Science and Technology (KAUST), Thuwal, Saudi Arabia} \\
 \textit{${^2}$Rain Neuromorphics, San Francisco, CA, USA} \\
 ok@ieee.org, foudam@uci.edu, ahmed.eltawil@kaust.edu.sa, khaled.salama@kaust.edu.sa
 }
 \vspace{-0.15in}
 }

\maketitle

\begin{abstract}

Designing generalized in-memory computing (IMC) hardware that efficiently supports a variety of workloads requires extensive design space exploration, which is infeasible to perform manually. Optimizing hardware individually for each workload or solely for the largest workload often fails to yield the most efficient generalized solutions. To address this, we propose a joint hardware-workload optimization framework that identifies optimised IMC chip architecture parameters, enabling more efficient, workload-flexible hardware. 
We show that joint optimization achieves 36\%, 36\%, 20\%, and 69\% better energy-latency-area scores for VGG16, ResNet18, AlexNet, and MobileNetV3, respectively, compared to the separate architecture parameters search optimizing for a single largest workload.
Additionally, we quantify the performance trade-offs and losses of the resulting generalized IMC hardware compared to workload-specific IMC designs.

\end{abstract}

\begin{IEEEkeywords}
Design space exploration, in-memory computing, software-hardware co-design, RRAM, hardware optimization
\end{IEEEkeywords}

\section{Introduction}


In-memory computing (IMC) is one of the promising solutions to implement energy- and area-efficient neural network accelerators, essential for advancing artificial intelligence (AI) applications \cite{zhang2020neuro, sebastian2020memory, smagulova2023resistive,fouda2022memory, aguirre2024hardware, yantir2022hardware, jain2019neural}. To optimize IMC system design and implement truly efficient IMC chips, a comprehensive design approach across multiple architecture hierarchy levels is essential. This includes addressing device-level design parameters and non-idealities,  optimizing circuit-level components and IMC macros, consideration of architecture-level design with attention to data transmission across chip components, and enhancing  software-level and workload performance \cite{krestinskaya2023towards, zhang2020neuro}. 
As AI models and the number of IMC hardware parameters grow in complexity, manual optimization of software and hardware parameters becomes infeasible.
Consequently, software-hardware co-design methodologies, such as hardware-aware neural architecture search (HW-NAS) and hardware space exploration, are essential for developing efficient and optimized IMC hardware \cite{mei2021zigzag, krestinskaya2024neural, sekanina2021neural}.

\begin{figure*}[t!]
    \includegraphics[width=\textwidth]{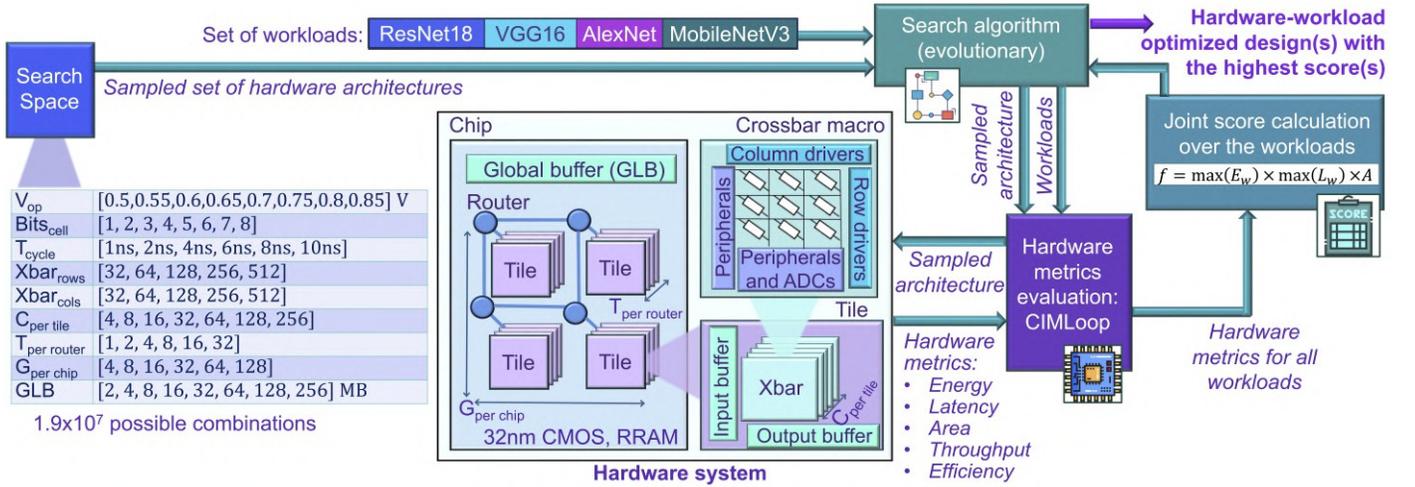}
    \vspace{-0.2cm}
    \caption{Joint hardware-workload optimization framework for IMC hardware.}
    \label{f1}
    \vspace{-0.35cm}
\end{figure*}

Current state-of-the-art methods and frameworks for design space exploration in IMC systems are primarily focused on optimizing of IMC hardware designs for specific workloads and applications \cite{negi2022nax, yang2021multi, sun2023gibbon, jiang2020device , moitra2023xpert, han2024comn}. However, IMC chips often need to be more generalized to support a variety of workloads effectively.
In addition, this field faces several open challenges, including the lack of unified framework to accommodate diverse AI models and different types of IMC hardware, runtime efficiency limitations and slow optimization speed, and insufficient architecture- and system-level considerations \cite{krestinskaya2024neural}. In this work, we address these challenges and propose high-speed joint hardware-workload design space exploration for IMC architectures, which aims to search for optimized generalized IMC system design parameters capable of supporting a range of different workloads effectively.



\begin{figure*}[th!]
    \includegraphics[width=\textwidth]{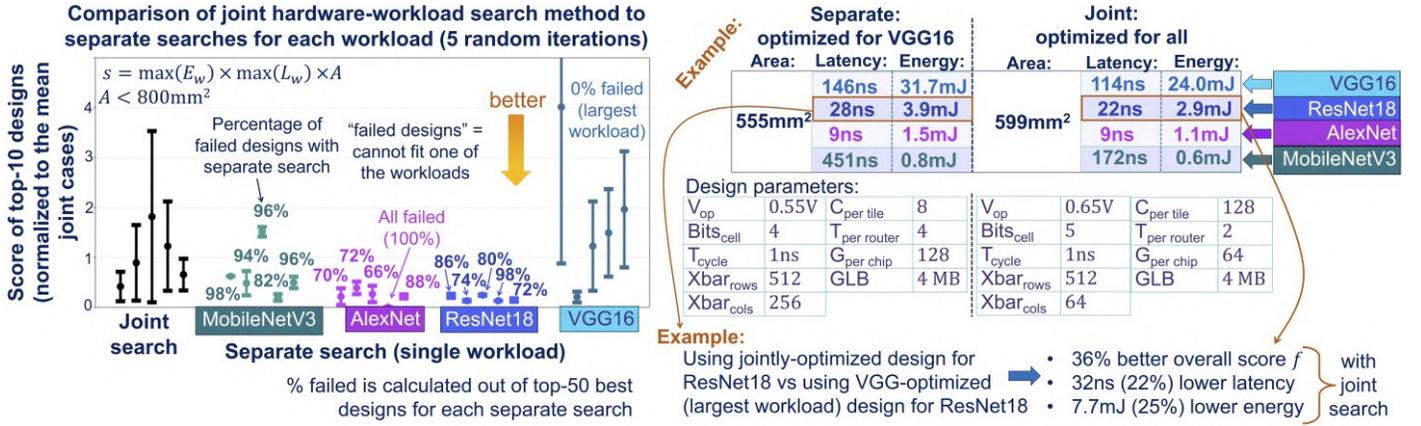}
    \vspace{-0.2cm}
    \caption{Importance of joint hardware-workload search for generalized architecture: performance comparison of top-10 designs across 5 random iterations for joint and separate searches, percentage of failed designs in separate search, and example contrasting separate optimization for the largest workload with joint optimization.}
    \label{f2}
        \vspace{-0.2in}
\end{figure*}

We demonstrate that hardware-workload co-optimization across multiple workloads is crucial for achieving optimal energy efficiency, on-chip area, and processing speed in a generalized IMC hardware system. Optimizing the design individually for each workload can result in 66–100\% of selected designs being incompatible with other workloads. Similarly, optimizing solely for the largest workload often yields suboptimal solutions. For instance, optimizing the IMC system for the VGG16 \cite{simonyan2014very} workload and then deploying it for ResNet18 \cite{he2016deep} results in lower performance compared to a joint optimization approach that considers all workloads. Specifically, joint optimization achieves a 36\% higher energy-area-latency score, 22\% lower latency, and 25\% lower energy consumption (see Fig. \ref{f2} in Section \ref{results}).




Overall, the main contributions of this work include implementing a joint hardware-workload framework to optimize generalized IMC hardware that supports multiple workloads, along with a comparative analysis against separate single-workload optimization. Additionally, our approach provides insights into the energy and latency trade-offs and performance loss involved when transitioning to a generalized design capable of supporting diverse workloads.


\section{State-of-the-art design space exploration for IMC hardware}
State-of-the-art methods for design space exploration in IMC hardware can be classified into three types: model parameters search for a fixed architecture, co-optimization of architecture and model, and hardware parameters optimization for a fixed model \cite{krestinskaya2024neural}. Model parameters search for a fixed IMC architecture focuses on optimizing neural network models to meet specific IMC hardware constraints, as demonstrated in \cite{benmeziane2023analognas, krestinskaya2020automating, yuan2021nas4rram, li2021flash, guan2022hardware}. This approach targets model optimization to align with IMC hardware limitations and constraints, maintaining high model accuracy while mitigating the effects of IMC device non-idealities on performance accuracy \cite{krestinskaya2020towards}. 
Co-optimization of model and architecture aims to identify an optimized pairs model and hardware parameters for specific applications, as seen in \cite{negi2022nax, sun2023gibbon, jiang2020device}. 
While hardware parameters optimization for a fixed model targets design space exploration to identify optimal hardware parameters that most effectively support a specific model \cite{yang2021multi, han2024comn}. 
A primary limitation of these frameworks is their focus on optimizing IMC hardware for specific neural network models, which restricts hardware generalizability. The output designs obtained from these frameworks lack the flexibility to accommodate diverse workloads. To address this gap, we propose a joint hardware-workload optimization framework aimed to search for optimized IMC hardware parameters that can effectively support multiple workloads.

\begin{table}[t!]
\caption{Comparison of the proposed framework with state-of-the-art approaches.}
\resizebox{\columnwidth}{!}{%
\begin{tabular}{|lcccccc|}
\hline
\multicolumn{1}{|c|}{\multirow{3}{*}{\textbf{\begin{tabular}[c]{@{}c@{}}Frame-\\ work\end{tabular}}}} & \multicolumn{1}{c|}{\multirow{3}{*}{\textbf{Algorihtm$^{*1}$}}}                 & \multicolumn{3}{c|}{\textbf{Optimization}}                                                                                                                   & \multicolumn{1}{c|}{\multirow{3}{*}{\textbf{\begin{tabular}[c]{@{}c@{}}Work-\\ load\end{tabular}}}} & \multirow{3}{*}{\textbf{\begin{tabular}[c]{@{}c@{}}Generalized \\ hardware \\ solution$^{*5}$\end{tabular}}} \\ \cline{3-5}
\multicolumn{1}{|c|}{}                                                                                & \multicolumn{1}{c|}{}                                                    & \multicolumn{1}{c|}{\multirow{2}{*}{\textbf{D$^{*2}$}}} & \multicolumn{1}{c|}{\multirow{2}{*}{\textbf{C$^{*3}$}}} & \multicolumn{1}{c|}{\multirow{2}{*}{\textbf{A$^{*4}$}}} & \multicolumn{1}{c|}{}                                                                               &                                                                                                       \\
\multicolumn{1}{|c|}{}                                                                                & \multicolumn{1}{c|}{}                                                    & \multicolumn{1}{c|}{}                              & \multicolumn{1}{c|}{}                              & \multicolumn{1}{c|}{}                              & \multicolumn{1}{c|}{}                                                                               &                                                                                                       \\ \hline
\multicolumn{1}{|l|}{\begin{tabular}[c]{@{}l@{}}NAX  \cite{negi2022nax}\end{tabular}}                          & \multicolumn{1}{c|}{DS}                                                  & \multicolumn{1}{c|}{-}                             & \multicolumn{1}{c|}{\checkmark}                             & \multicolumn{1}{c|}{-}                             & \multicolumn{1}{c|}{single}                                                                         & -                                                                                                     \\ \hline
\multicolumn{1}{|l|}{\begin{tabular}[c]{@{}l@{}}Yang X. \\ et. al \cite{yang2021multi}\end{tabular}}               & \multicolumn{1}{c|}{\begin{tabular}[c]{@{}c@{}}CF-\\ MESMO\end{tabular}} & \multicolumn{1}{c|}{\checkmark}                             & \multicolumn{1}{c|}{\checkmark}                             & \multicolumn{1}{c|}{-}                             & \multicolumn{1}{c|}{single}                                                                         & -                                                                                                     \\ \hline
\multicolumn{1}{|l|}{\begin{tabular}[c]{@{}l@{}}Gibbon  \cite{sun2023gibbon}\end{tabular}}                       & \multicolumn{1}{c|}{EA}                                                  & \multicolumn{1}{c|}{\checkmark}                             & \multicolumn{1}{c|}{\checkmark}                             & \multicolumn{1}{c|}{-}                             & \multicolumn{1}{c|}{single}                                                                         & -                                                                                                     \\ \hline
\multicolumn{1}{|l|}{\begin{tabular}[c]{@{}l@{}}NACIM  \cite{jiang2020device}\end{tabular}}                        & \multicolumn{1}{c|}{RL}                                                  & \multicolumn{1}{c|}{\checkmark}                             & \multicolumn{1}{c|}{-}                             & \multicolumn{1}{c|}{\checkmark}                             & \multicolumn{1}{c|}{single}                                                                         & -                                                                                                     \\ \hline
\multicolumn{1}{|l|}{\begin{tabular}[c]{@{}l@{}}XPert \cite{moitra2023xpert}\end{tabular}}                          & \multicolumn{1}{c|}{DS}                                                  & \multicolumn{1}{c|}{-}                             & \multicolumn{1}{c|}{\checkmark}                             & \multicolumn{1}{c|}{-}                             & \multicolumn{1}{c|}{single}                                                                         & -                                                                                                     \\ \hline
\multicolumn{1}{|l|}{\begin{tabular}[c]{@{}l@{}}CoMN \cite{han2024comn}\end{tabular}}                           & \multicolumn{1}{c|}{BO}                                                  & \multicolumn{1}{c|}{\checkmark}                             & \multicolumn{1}{c|}{\checkmark}                             & \multicolumn{1}{c|}{\checkmark}                             & \multicolumn{1}{c|}{single}                                                                         & -                                                                                                     \\ \hline
\multicolumn{1}{|l|}{\textbf{Ours}}                                                                            & \multicolumn{1}{c|}{EA}                                                  & \multicolumn{1}{c|}{\checkmark}                             & \multicolumn{1}{c|}{\checkmark}                             & \multicolumn{1}{c|}{\checkmark}                             & \multicolumn{1}{c|}{\textbf{multiple}}                                                                          & \checkmark                                                                                                     \\ \hline
\multicolumn{7}{|l|}{\begin{tabular}[c]{@{}l@{}}$^{*1}$: DS - differential search, CF-MESMO - continuous-fidelity max-value \\ entropy search for multi-objective optimization, EA - evolutionary algorithm, \\RL - reinforcement learning, BO - Bayesian optimization.
\\$^{*2}$: \textbf{Device optimization} - considering optimization of the device  parameters, \\ e.g. bits per cell. $^{*3}$: \textbf{Circuit optimization} - considering  optimization of \\ circuit-level parameters of crossbar macros,  e.g. array size or ADC precision.  \\  $^{*4}$: \textbf{Architecture optimization} -  consideration of higher level architecture \\parameters, e.g. tiles, connections between them, global buffer, etc.\\ $^{*5}$: Generalized hardware solution output supporting different workloads\end{tabular}}             \\ \hline
\end{tabular}
}
\label{t1}
\vspace{-0.5cm}
\end{table}

Table \ref{t1} provides a comparison between the proposed method and state-of-the-art approaches.
Most state-of-the-art frameworks adopt a two-stage optimization strategy, where a specific neural network model is optimized first, followed by hardware optimization tailored to this workload. For instance, XPert framework \cite{moitra2023xpert} uses a two-stage optimization approach by initially optimizing the VGG16 \cite{simonyan2014very} channel depth, ADC parameters, and column sharing configurations and subsequently performing ADC and input precision optimization. However, in practical scenarios, hardware chips are often required to support multiple, varied workloads rather than being optimized for a single, fixed task.

Moreover, most hardware design exploration frameworks focus on device-level parameters, e.g. optimal bits per cell, as well as circuit parameters related to the IMC crossbar macro and peripheral circuits, including crossbar size, ADC precision, and number of ADCs per macro \cite{negi2022nax, yang2021multi, sun2023gibbon, moitra2023xpert}. 
However, to achieve a realistic assessment of IMC chip performance, it is essential to consider higher-level architectural hierarchies. These often contribute significantly to energy consumption and area overhead, so effective energy efficiency of IMC system can be 100 times lower than the performance measured at the macro level \cite{jain2022heterogeneous}. A recent framework focusing on larger space of hardware parameters is CoMN \cite{han2024comn}, which considers optimization architecture-related parameters, e.g. tile sizes and buffer sizes, for a specific workload.
In this work, we also consider optimization of architecture hierarchy parameters, but optimizing for a more generalized case, accommodating diverse workloads.

\section{Joint hardware-workload optimization}

\subsection{Framework overview}

We propose a joint hardware-workload optimization approach and develop the framework searching for optimized IMC-based chip architecture parameters that support different workloads, resulting in a more generalized hardware solution, as shown in Fig. \ref{f1}.
The framework takes as inputs a set of workloads and a defined search space, and outputs hardware-workload-optimized designs that achieve the highest score in the specified objective function, including optimized hardware parameters and corresponding performance metrics. Experiments are conducted with different objective functions (see Section \ref{results}). The framework performs IMC hardware optimization by jointly considering all workloads, using the highest metrics across workloads to calculate the objective function. Compared to sequential optimization or optimizing solely for the largest workload, this approach yields a more generalized hardware solution, effectively optimized to support and perform well across all specified workloads.



\subsection{Hardware Performance Evaluation}

In this work, we conduct the experiments considering several levels of IMC hardware hierarchy shown in Fig. \ref{f1}. We simulate a tiled, crossbar-based architecture with resistive random access memory (RRAM) devices (from \cite{lu2021neurosim}) and 32nm CMOS technology. 
 Each tile consists of $\mathrm{C_{per \, tile}}$ crossbars along with peripheral circuits, ADCs, column and row drivers, and input/output buffers. 
 Data transmission between tiles occurs through shared routers, following the architecture in \cite{shafiee2016isaac}, where each router connects to $\mathrm{T_{per \, router}}$ tiles. The overall chip structure includes $\mathrm{G_{per \, chip}}$ tile groups and a global buffer.

 For hardware estimations, we use CIMLoop \cite{andrulis2024cimloop}, which integrates Timeloop \cite{parashar2019timeloop} for mapping and employs Accelergy \cite{wu2019accelergy} for energy estimations. 
The crossbar-related estimations in CIMLoop are equivalent to those in NeuroSim \cite{peng2020dnn+}, but with reduced simulation time, which significantly decreases the search duration for optimized hardware configurations.

The search space, shown in Fig. \ref{f1}, includes crossbar sizes (both rows $\mathrm{Xbar_{rows}}$ and columns $\mathrm{Xbar_{cols}}$), number of crossbars per tile $\mathrm{C_{per \, tile}}$, number of tiles per router $\mathrm{T_{per \, router}}$, number of tile groups within the chip $\mathrm{G_{per \, chip}}$. Additionally, we include operating voltage $\mathrm{V_{op}}$, impacting both energy efficiency and throughput \cite{andrulis2024cimloop}. Moreover, we consider the number of bits in RRAM cell $\mathrm{Bits_{cell}}$ (which affects hardware configuration rather than performance accuracy), cycle time $\mathrm{T_{cycle}}$ (representing operating frequency), and the size of the global buffer $\mathrm{GLB}$ storing input and output data. Overall, this search space contains approximately $1.9\times10^7$ configurations.

\subsection{Algorithm Selection}


The most commonly used optimization algorithms for model and hardware design space exploration (Table \ref{t1}) include evolutionary algorithms (EA), differential search (DS), reinforcement learning (RL), and Bayesian optimization (BO) \cite{krestinskaya2024neural}. DS is particularly well-suited for differentiable search spaces in HW-NAS with constrained optimization, where only software model parameters are optimized under hardware constraints or co-optimization of software and hardware parameters, as demonstrated in \cite{moitra2023xpert}. In this work, we use an evolutionary algorithm, as it performs faster than RL and BO for a discrete search space of moderate size \cite{krestinskaya2024neural}. 

The evolutionary algorithm adopted for this work is a genetic algorithm of $G$ generations and population size $P$
\cite{blank2020pymoo}. The initial population is randomly selected from the architecture search space. However, any configuration that fails to accommodate the largest workload in the set is discarded, ensuring that the initial population consists only of valid solutions. After sampling initial population, the hardware metrics of these architectures are evaluated and sorted based on the objective function (score) calculated as 
\begin{equation}\label{eq:o2}
\begin{aligned}
f (E_w, L_w, A),  \,  
\mathrm{s.t.} \, A \leq A_{constr}
\end{aligned}
\end{equation}
 where $E_w$ is the energy required to process a specific workload on the sampled IMC hardware, $L_w$ is the latency of this workload, $A$ and $A_{constr}$ represent the on-chip area of this hardware and area constraint, respectively. For example, the objective function can be calculated as $f=\max(E_w)\times \max(L_w) \times A$, where for each generation, we consider the highest latency $\max(L_w)$ and energy $\max(E_w)$ across all workloads in the objective functio aiming to minimize these scores. In Section \ref{results}, we also demonstrate the experiments with the other constraint objective functions.

After evaluating the samples, the crossover operation is performed to generate a new set of architectures by combining parameters from the previous population. This is followed by the mutation phase, where the newly constructed architectures undergo further modifications. This work employs simulated binary crossover and polynomial mutation \cite{deb2007self, blank2020pymoo}, with a crossover probability of 0.95 and a distribution index of 3 (within the typical range of 3 to 30). These settings prioritize exploration, promoting diversity in the new population and ensuring broad coverage of the search space.
Throughout the process, we store the performance scores and hardware metrics of all sampled architectures after each generation. After the mutation phase, the evaluation step is repeated for $G$ generations. The best set of architectures is then selected from the stored population history.




\begin{figure}[t]
    \includegraphics[width=\columnwidth]{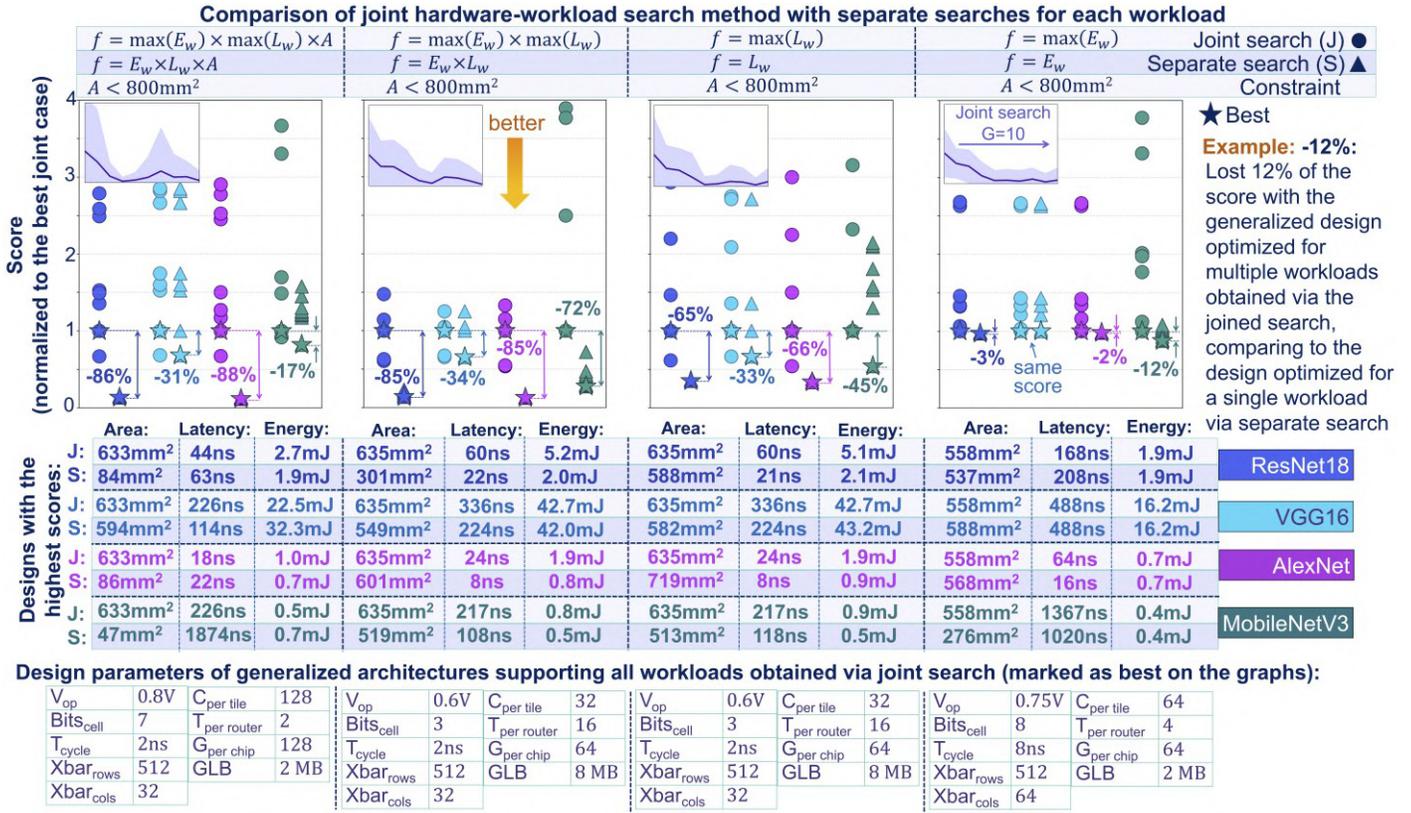}
    \caption{Comparison of joint and separate searches with analysis of score loss in transition to generalized designs.}
    \label{f3}
    \vspace{-0.15in}
\end{figure}

\section{Experimental results}
\label{results}

The workloads evaluated in these experiments cover the entire spectrum of CNN model types, including ResNet18 \cite{he2016deep}, VGG16 \cite{simonyan2014very}, AlexNet \cite{krizhevsky2012imagenet} and MobileNetV3 \cite{howard2019searching}. We apply 8-bit quantization to both the model's weights and inputs, with the ADC precision also fixed at 8 bits. Each joint optimization search (with the population size $P=40$ running for $G=10$ generations) takes approximately 4 hours on an AMD processor with 64 cores. 
Across experiments with both constrained and unconstrained objective functions, we observed that unconstrained searches often led to designs with excessively large on-chip area (Fig. \ref{f1}). Thus, adding area constraints is crucial, and we present the results for constrained searches accordingly. In this section, we refer to the proposed optimization as 'joint search' and to optimization over individual workloads separately as 'separate search'.



Figure \ref{f2} demonstrates the performance of the joint search (top-10 best selected designs) across 5 random iterations with different randomly selected initial populations, compared to separate searches over individual workloads. The final scores of separate searches are recalculated for fair comparison with joint optimization for the performance across all workloads. The graph indicates that, in separate optimization for specific workloads, most of the best-selected designs fail to support all workloads ('failed designs' in Fig. \ref{f2}), except for the largest workload, VGG16. While hardware optimized solely for the largest workload performs worse than joint optimization, as evidenced by implementing ResNet18 on VGG16-optimized hardware versus a jointly optimized hardware (Fig. \ref{f2}, on the right). Overall, the example in Fig. \ref{f2} demonstrates that joint optimization achieves 36\%, 36\%, 20\%, and 69\% better energy-latency-area scores for VGG16, ResNet18, AlexNet, and MobileNetV3, respectively, compared to the separate search optimizing for the single largest workload (VGG16).


When implementing a generalized architecture supporting multiple workloads, some performance metrics, such as energy or latency, are inevitably sacrificed compared to a design optimized for a single workload. Fig. \ref{f3} illustrates the performance loss in the generalized IMC architecture obtained via joint optimization, comparing it to workload-specific architectures and showing the percentage loss in score as the architecture becomes more generalized. Each graph presents the top-10 designs from a joint hardware-workload search alongside results from separate searches optimized individually for four different workloads under a specific objective function.
Scores in each graph are normalized to the best architecture from the joint search, and convergence curves over 10 generations of the evolutionary algorithm are provided for each joint search case. To ensure a fair comparison, each search begins with the same set of initial architectures (using a specified random seed). 
Overall, the generalized models can lose from 17\% to 86\% (Fig. \ref{f3}) of the score comparing to workload-specific designs.
Additionally, we highlight the top designs selected in each joint search scenario.

\section{Conclusion}

We proposed the joint hardware-workload optimization framework for IMC design space exploration, aimed at identifying an optimized, generalized IMC hardware solution capable of supporting a variety of workloads. 
Our results demonstrate that this joint optimization approach leads to optimized designs with better performance compared to optimizing hardware for each workload individually or solely for the largest workload. Additionally, we analyzed the performance trade-offs and losses involved in transitioning from IMC solutions optimized for specific applications to more generalized solutions capable of supporting diverse workloads.




\section*{Acknowledgement}
This work has been partially supported by King Abdullah University of Science and Technology CRG program under grant number:  URF/1/4704-01-01


\end{document}